\documentclass[prb,twocolumn,showpacs]{revtex4}
\usepackage{graphicx}
\usepackage{amssymb}
\usepackage{dcolumn}
\usepackage{float}
\usepackage{bm}
\usepackage{booktabs}
\usepackage{amsmath} 
\usepackage[utf8]{inputenc}
\usepackage{lmodern}
\usepackage{booktabs}
\usepackage{color}
\usepackage{color}
\usepackage[breaklinks=true,colorlinks,citecolor=magenta,linkcolor=blue,urlcolor=magenta]{hyperref}

\newcommand{\nn}{\nonumber}

\newcolumntype{M}[1]{>{\centering\arraybackslash}m{#1}}

\newcommand{\bs}{\boldsymbol}
\DeclareMathAlphabet{\bi}{OML}{cmm}{b}{it}

\def\be{\begin{equation}}
\def\ee{\end{equation}}
\def\bearr{\begin{eqnarray}}
\def\eearr{\end{eqnarray}}
\def\la{\langle}
\def\ra{\rangle}

\begin{document}


\title{Thermoelectric and optical probes for a Fermi surface topology change
in noncentrosymmetric metals}

\bigskip
\author{Sonu Verma,${}^1$ 
Tutul Biswas,${}^2$ and Tarun Kanti Ghosh${}^1$\\
\normalsize
$\color{blue}{^1}$ Department of Physics, Indian Institute of Technology-Kanpur,
Kanpur-208 016, India\\
$\color{blue}{^2}$  Department of Physics, 
University of North Bengal, Raja Rammohunpur-734 013, India}

\date{\today}

\begin{abstract}
Noncentrosymmetric metals such as Li$_2$(Pd$_{1-x}$Pt$_x$)$_3$B have different Fermi surface topology 
below and above the band touching point where spin-degeneracy is not lifted by the spin-orbit coupling. 
We investigate  thermoelectric and optical response as probes for this Fermi surface topology change.
We show that the chemical potential displays a dimensional crossover from a three-dimensional to 
one-dimensional characteristics as the descending Fermi energy crosses the band touching point. 
This dimensional crossover is due to the existence of different Fermi surface topology 
above and below the band touching point. 
We obtain an exact expression of relaxation time due to short-range scatterer by 
solving Boltzmann transport equations self-consistently. 
The thermoelctric power and figure of merit are significantly enhanced as 
the Fermi energy goes below the band touching point owing to the underlying one-dimensional-like nature of noncentrosymmteric bulk metals. 
The value of thermoelectric figure of merit goes beyond two
as the Fermi energy approaches to the van Hove singularity for lower spin-orbit coupling.   
Similarly, the studies of the zero-frequency and finite-frequency
optical conductivities in the zero-momentum limit reflect the 
nature of topological change of the Fermi surface.
The Hall coefficient and optical absorption width exhibit distinct signatures 
in response to the changes in Fermi surface topology. 

%
\end{abstract}


\maketitle

\section{Introduction}
Breaking of inversion symmetry has far reaching consequences in 
condensed matter physics. It gives rise to spin-orbit interaction (SOI), 
which itself serves as the backbone of the rich fast-growing field of 
spintronics \cite{spintro1,spintro2, spintro3}. At the interface of 
semiconductor heterostructures, the inversion symmetry is broken due to 
the band mismatch/external electric field, 
thus giving rise to a particular type of SOI known as Rashba spin-orbit 
interaction (RSOI) \cite{Rash1,Rash2}. Besides the breaking of inversion 
symmetry in bulk semiconductors having zinc blende structures causes 
Dresselhaus spin-orbit interaction \cite{Dress}. 
The RSOI has potential applications in developing spintronic devices as its strength 
is externally tunable \cite{Nitta} and therefore it is mostly studied. It is revealed 
that the RSOI can host a plethora of exotic phenomena such as dissipationless spin 
current \cite{dissp1, dissp2}, spin Hall effect \cite{SHE1,SHE2,SHE3,SHE4,SHE5,SHE6}, 
spin-orbit torque \cite{SHE6,spin-torque} and spin galvanic effect \cite{SHE6,SGE1}. 

Angle resolved photo emission spectroscopy has confirmed the existence 
of large spin splitting in several systems such as Bi/Ag(111) surface 
alloy \cite{Bi_Alloy}, topological insulators like Bi$_2$Se$_3$ etc. \cite{BiSe1,BiSe2}. 
The spin-orbit coupling strength found in these systems are larger in magnitude 
at least by two orders than that found in conventional semiconductor heterostructures.
The recent discovery of giant RSOI \cite{BiTeI1,BiTeI2,BiTeI3,BiTeI4,BiTeI5} in three-dimensional (3D) polar semiconductor BiTeX (X=Cl, Br, I) has triggered immense
investigations in the field of spintronics both theoretically and experimentally. 
The surface states of BiTeX exhibit large Rashba splitting as a result of surface-induced asymmetry. The origin of giant RSOI in the bulk of such materials
has been unveiled by ${\bf k}\cdot{\bf p}$ perturbation analysis \cite{BiTeI2} and 
is attributed to the distinct crystal structure of BiTeX 
and large SOI of Bi. Itinerant electrons also experience strong RSOI in
B20 compounds \cite{B201} and noncentrosymmetric metals such as 
Li$_2$(Pd$_{1-x}$Pt$_x$)$_3$B\cite{NC_Li1}. The material Li$_2$(Pd$_{1-x}$Pt$_x$)$_3$B 
exhibits superconductivity \cite{sup1,rev-sup} as result of inversion symmetry breaking, whereas
B20 compounds \cite{B202,B203,B204,B205}
host nontrivial spin textures including spin helix and magnetic skyrmions. 
The spin-momentum locking in Li$_2$(Pd$_{1-x}$Pt$_x$)$_3$B has distinct structure 
than that in the BiTeX family. Therefore, this noncentrosymmetric material has drawn 
immense interest from the perspective of superconductivity \cite{NC_Li2}, Kerr rotation \cite{NC_Li1,NC_Li3}, spin susceptibility, \cite{spin_scp} and
Ruderman-Kittel-Kasuya-Yosida interaction \cite{RKKY}.

To probe electronic states in BiTeX materials various investigations have been 
performed recently in the context of transport \cite{Trans1,Trans2,Trans3,Trans4,Trans5,Trans6,Trans7}, 
thermoelectric \cite{Therm1,Therm2}, and optical \cite{Trans3,Mag_pht,Opt3,op-cond-ti}
response. In addition there have also been theoretical and experimental studies \cite{susRashba,susprl} 
where magnetic susceptibility of these systems changes its nature from paramagnetic to diamagnetic as 
Fermi energy crosses the band touching point from below. However, such studies are still 
missing in the case of spin-orbit coupled noncentrosymmetric metal like Li$_2$(Pd$_{1-x}$Pt$_x$)$_3$B,  
which has different Fermi surface topology below and above the band touching point. This lack 
of information motivates us to address the issue that whether the thermoelectric and 
optical probes can be used to extract the information about the topology of the 
Fermi surface in noncentrosymmetric metals like Li$_2$(Pd$_{1-x}$Pt$_x$)$_3$B.

Transport properties of spin-orbit coupled condensed matter systems 
at low temperature is of great interest for various reasons. 
It is of primary interest to reduce the thermal conductivity  to obtain high 
thermoelectric figure of merit being inversely proportional to the thermal conductivity.
The thermal energy is transported by the electrons as well as phonons.
At temperatures much smaller than the Debye temperature $\theta_D$,
number of phonons participating in thermal transport will be very small and electronic thermal
conductivity will dominate over the lattice counterparts.
For instance, in B20 compounds like Fe$_{1-x}$Co$_{x}$Si \cite{FeSi-debye1} 
$\theta_D$ is around  350 K and in MnGe, CoGe \cite{MnGe-debye2} $\theta_D$ is about (269--281) K.
Therefore, in this work we consider the temperature around (5--20) K which is much smaller
than $\theta_D$. 
Moreover, it is essential to have low temperature of the system
so that thermal energy is always less than the spin splitting energy which is required
to control spin of a charge carrier for spintronic device applications.


In this work, we provide a systematic study of thermoelectric transport 
coefficients and optical responses in noncentrosymmetric metals. We find that the 
chemical potential exhibits a dimensional crossover from a 3D- to 1D-like characteristics as 
the Fermi energy goes below the band touching point (BTP). This feature is attributed 
to the existence of different Fermi surface topology above and below the BTP. 
We obtain exact expression of relaxation time assuming short-range electron-impurity
scattering, 
by solving the Boltzmann transport equations including interband scattering self-consistently. 
We provide results of all thermoelectric transport coefficients.  
The thermoelectric power and figure of merit are significantly enhanced below BTP owing to 
the underlying 1D-like nature of this system as a consequence of change in Fermi surface topology.  
We obtain a remarkable value, more than 2 of thermoelectric 
figure of merit for  $\alpha=0.2~$eV-nm at low density below BTP.
Similarly, the studies of the zero-frequency and finite-frequency 
optical conductivities shed some light on the nature of spin-split energy gap 
and would help to extract the spin-orbit coupling strength.
We find that Hall coefficient and optical absorption width respond differently 
to the change in the Fermi surface topology.




This paper is organized as follows. In Sec. II, we provide a discussion on the 
ground state properties of the physical system considered.
In Sec. III, we discuss various thermoelectric properties. Sec. IV includes 
information of the Drude 
weight, Hall coefficient, and finite-frequency optical conductivity.
We summarize our main results in Sec. V.

\section{Ground state properties}
\begin{figure}[htbp]
\begin{center}\leavevmode
\includegraphics[width=0.45\textwidth]{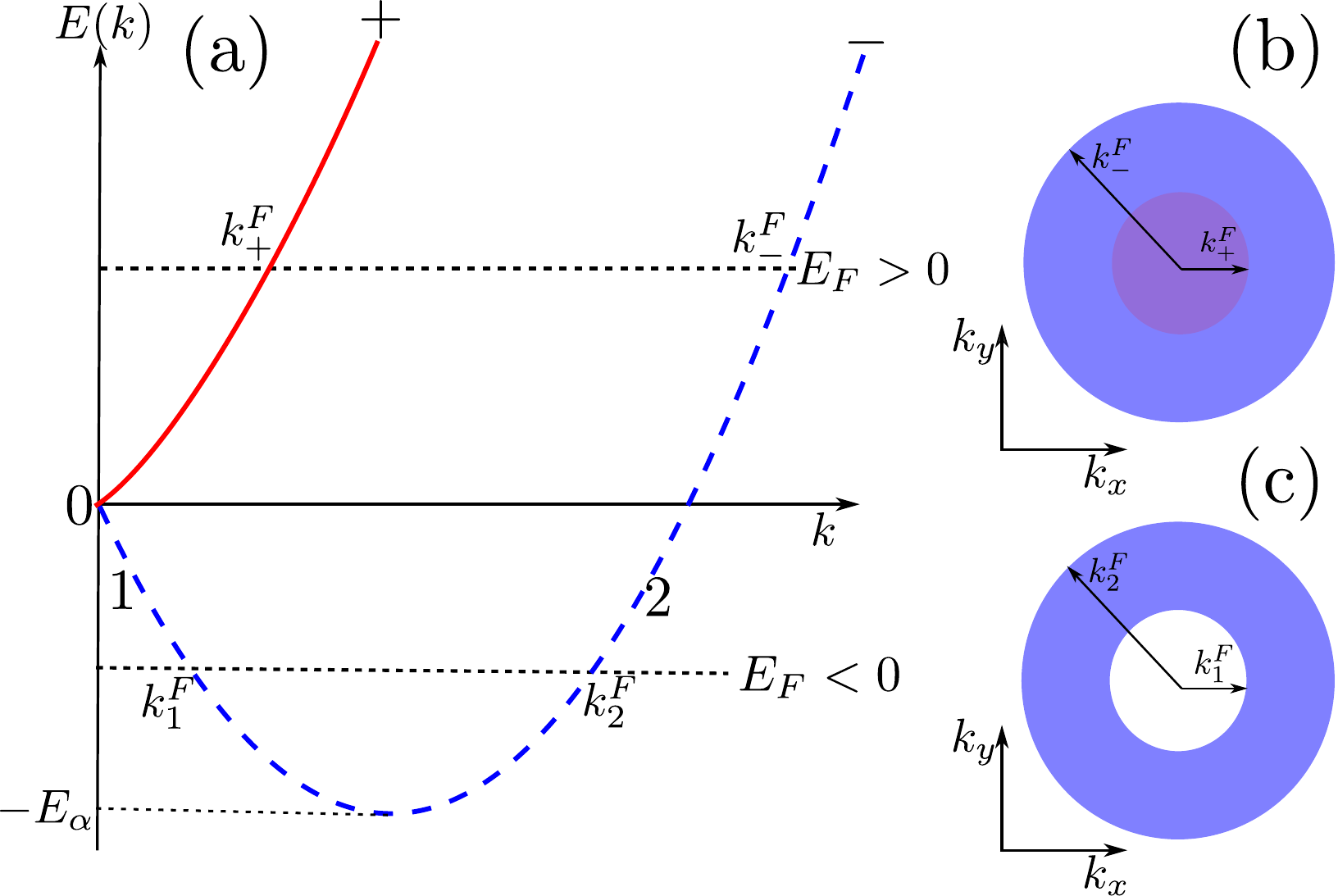}
\caption{(a) Spin-split energy bands of the noncentrosymmetric metals:
The two bands touch at $k=0$ which is known as band touching point (BTP).
(b) Cross-sections of the Fermi surfaces for $E_F > 0$ and $E_F < 0$.
The Fermi surface topology has convex-convex shape  and concave-convex shape for $E_F>0$ and  
$E_F<0$, respectively. The Fermi surface topology changes at band touching point (at $E_F=0$).}
\label{sketch}
\end{center}
\end{figure}

We consider conduction electrons in a 3D noncentrosymmetric metal. 
As mentioned in the introduction, the usual examples 
of noncentrosymmetric metals are Li$_2$(Pd$_{3-x}$Pt$_x$)B and B20 compounds. 
All these materials possess cubic crystal structure. In this particular lattice 
geometry, the low-energy conduction electrons can be effectively described by 
the following Hamiltonian based on symmetry analysis \cite{B201,NC_Li2}:
\begin{eqnarray}\label{Ham1}
H=\frac{\hbar^2 {\bf k}^2}{2m^\ast} \sigma_0 + \alpha  \; {\bs \sigma} \cdot {\bf k},
\end{eqnarray}
where $m^\ast$ is the effective mass of electron, $\sigma_0$ is 
$2 \times 2 $ identity matrix, ${\bs \sigma} = \{\sigma_x,\sigma_y,\sigma_z \}$ 
is Pauli spin matrix, $ {\bf k} = \{k_x, k_y, k_z\}$ with 
$ k= \sqrt{k_x^2+k_y^2+k_z^2}$ is the electron's wavevector, and $\alpha $ is 
the strength of the RSOI. 
The RSOI term in Eq. (\ref{Ham1}) has the form ${\bs \sigma}\cdot{\bf k}$ in which 
$k_x$, $k_y$, and $k_z$ are intertwined with $\sigma_x$, $\sigma_y$, and $\sigma_z$, 
respectively. This distinct spin-momentum locking is absent in 2D Rashba systems 
and 3D Rashba semiconductors such as BiTeX and therefore gives rise to particular 
Fermi surface topology, different than other Rashba systems, which will be 
discussed shortly. Note that the Hamiltonian commutes with the helicity operator 
${\bf k}\cdot{\bs \sigma}/k$
so that its eigenvalues $\pm 1$ are good quantum numbers. Thus, the eigenstates of the 
above Hamiltonian can be obtained as eigenstates of the helicity operator modulated 
by a plane wave like 
$\psi_{{\bf k}}^\lambda({\bf r})=\phi_{\bf k}^\lambda e^{i {\bf k \cdot r}}/\sqrt{V}$, 
where $V$ is volume of the system, $\lambda = \pm$ represents two opposite helicity, 
and $\phi_{\bf k}^\lambda$ is helicity eigenstate which takes the following forms:
$\phi_{\bf k}^+=[\cos(\theta/2)~~e^{i\phi} \sin(\theta/2)]^\mathcal{T}$ for $\lambda = +$ and 
$\phi_{\bf k}^-=[\sin(\theta/2)~~-e^{i\phi}\cos(\theta/2)]^\mathcal{T}$ for $\lambda = -$,
with $\mathcal{T}$ being the transpose operation. 
Here, $\theta$ and $\phi$ are the polar and azimuthal angle, respectively, which represent 
the orientation of ${\bf k}$. The energy dispersion consists of two spin-split bands 
corresponding to $\lambda = \pm $ having the structure 
$E_{{\bf k}\lambda} = \hbar^2 k^2/(2m^\ast) + \lambda \alpha k$. This dispersion is 
depicted in Fig. \ref{sketch}(a).
The full bandstructure calculations in Li$_2$Pd$_3$B \cite{fullbandstruc}
	shows that the low-energy bands around the $\Gamma$  point with spin-orbit coupling are similar 
        to Fig. \ref{sketch}(a). Comparing with the band structure calculations, 
	the model Hamiltonian appears to be valid for $- 30$ meV $ < E < 0.1 $ eV 
	with a typical value of $\alpha = 0.1$ eV nm.

The band $E_{{\bf k}-}$ has a nonmonotonous behavior for 
$E<0$ and attains a minimum value 
$E_{\rm min} = - E_\alpha = - \hbar^2 k_\alpha^2/(2m^\ast)$ at 
$k=k_\alpha = m^\ast\alpha/\hbar^2$.
In Figs. \ref{sketch}(b) and \ref{sketch}(c), constant
energy surfaces are shown for $E>0$ and $E<0$, respectively. The wavevectors 
corresponding to $E>0$ are given by 
$k_{\lambda} = -\lambda k_{\alpha} + \sqrt{k_{\alpha}^2 + 2m^\ast E/\hbar^2}$. 
Here, $k_{\pm}$ represent the radii of the two concentric spherical constant energy
surfaces as shown in Fig. \ref{sketch}(b). For $E>0$, the topology of the Fermi surface 
has convex-convex shape on $\lambda=+$ and $\lambda=-$ bands, respectively. The corresponding 
density of states  for 
$\lambda = \pm $ bands are given as
\begin{eqnarray}\label{dof1}
D_{\lambda}^{>}(E) & = &  \frac{1}{4\pi^2}\Big(\frac{2m^\ast}{\hbar^2}\Big)^{\frac{3}{2}}
\Bigg[\frac{E+2E_\alpha}{\sqrt{E+E_\alpha} }
-\lambda \sqrt{4E_\alpha} \Bigg].
\end{eqnarray}
On contrary for $ E<0$, there exists only one energy band $E_{\bf k}^-$
and the topology of energy surface changes completely as compared to $E>0$. 
For $E<0$, the topology of the Fermi surface has concave-convex shape on the inner 
and outer branches, respectively. The cross-section of the Fermi surface for $E<0$ is shown in Fig. \ref{sketch}(c). 
This is characterized by the following wave vectors
$ k_{\eta} = k_{\alpha} -(-1)^{\eta-1} \sqrt{k_{\alpha}^2 + 2m^\ast E/\hbar^2}$, $\eta=1,2$. It is worthy 
to mention that $ 0\leq k_1 \leq k_{\alpha} $ and 
$ k_{\alpha} \leq k_2 \leq 2  k_{\alpha} $.
The region between two concentric spherical shells with 
the inner radius $k_1$ and outer radius $k_2$ contains the following available 
density of states
\begin{eqnarray}\label{dof2}
D_{\eta}^{<}(E) = \frac{1}{4\pi^2}\Big(\frac{2m^\ast}{\hbar^2}\Big)^{\frac{3}{2}}
\Bigg[\frac{E+2E_\alpha}{\sqrt{E+E_\alpha} }
-(-1)^{\eta-1} \sqrt{4E_\alpha} \Bigg].
\end{eqnarray}

It is interesting to note that there is an inverse square-root divergence of 
$D_{\eta}^{<}(E)$  as $E \rightarrow - E_{\alpha}$, similar to 
the van Hove singularity in conventional one-dimensional systems as well as 
in 2D Rashba systems. The large DOS in the very low-density 
($E\rightarrow - E_{\alpha}$) limit is due to the nonvanishing spherical 
energy surfaces with the radii $k_{1,2}$ approach to $ k_{\alpha}$, and vanishing
velocity as ${\bf v}({\bf k}) = (1/\hbar) {\bs \nabla}_{\bf k} E_{{\bf k}-}
\propto \sqrt{E + E_{\alpha}}$.

\begin{figure}[htbp]
\begin{center}\leavevmode
\includegraphics[width=0.45\textwidth]{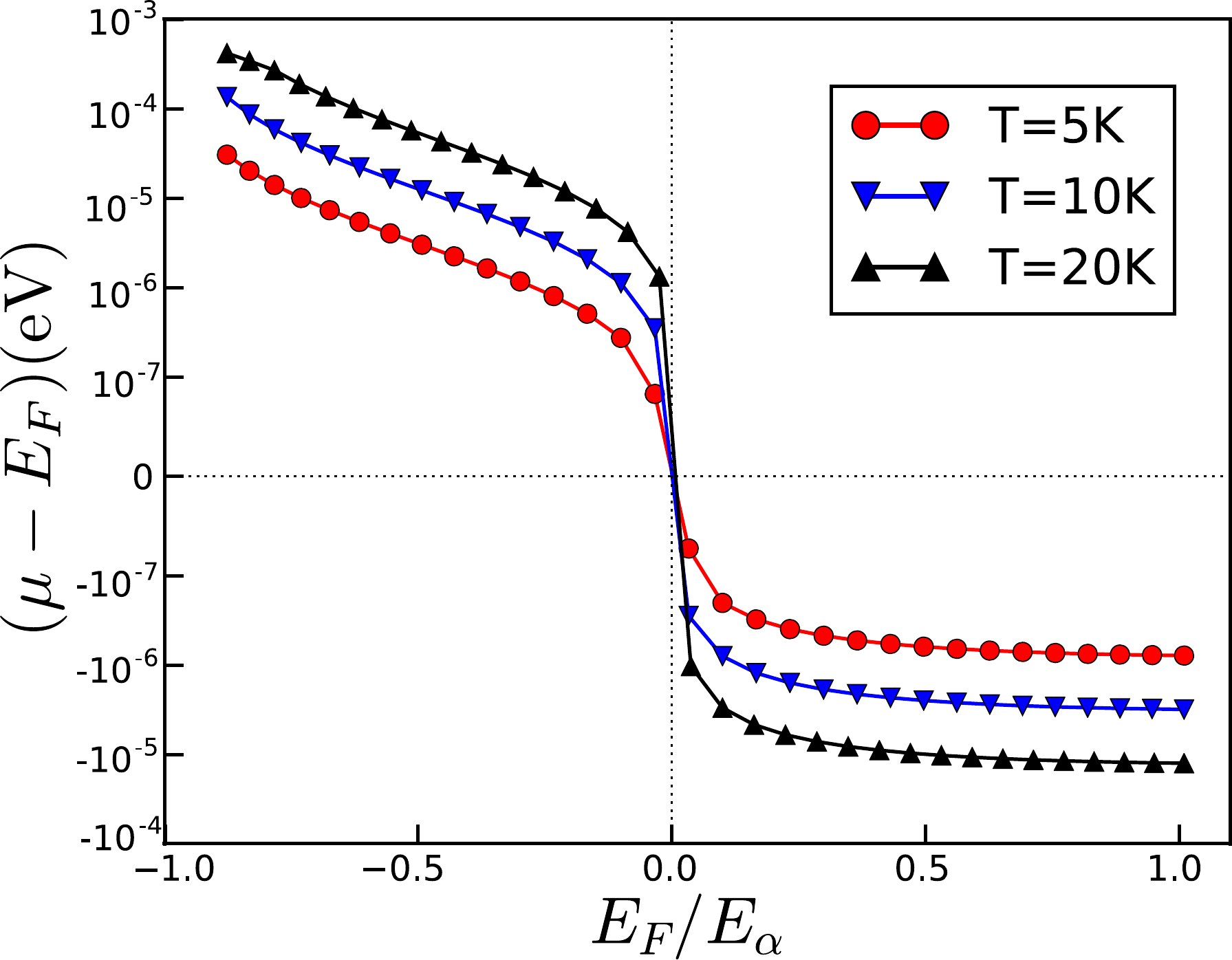}
\caption{Difference between $\mu$ and $E_F$ as a function of $E_F$ at 
a fixed $\alpha = 10^{-10}$ eV m for $T=5$, $10$, and $20$ K.
The change in sign of $\mu - E_F$ at $E_F=0$ implies the dimensional
crossover.} 
\label{chemical}
\end{center}
\end{figure}

For a given electron density $n_e$, the chemical potential $\mu$ at temperature $T$
can be obtained from the normalization condition,
\begin{small}
\begin{eqnarray}\label{Norm_C}
n_e = \sum_{\eta} \int_{- E_{\alpha}}^{0} f(E_\eta) D_{\eta}^{<}(E) dE 
+  \sum_{\lambda} \int_{0}^{\infty} f(E_\lambda) D_{\lambda}^{>} (E) dE,
\end{eqnarray}
\end{small}
where $f(E) = [e^{(E-\mu)/(k_BT)} + 1 ]^{-1}$ is the Fermi-Dirac distribution 
function.
In the $T \rightarrow 0$ limit, we obtain Eq. (\ref{Norm_C}) in the following form 
to extract Fermi energy $E_F$
\begin{eqnarray}\label{fermi}
3\pi^2 n_e & = & \sqrt{k_{\alpha}^2 + \frac{2m^\ast E_F}{\hbar^2}} 
\Big[4 k_{\alpha}^2 + \frac{2m^\ast E_F}{\hbar^2} \Big]. 
\end{eqnarray} 
Note that with $\alpha = 0$ Eq. (\ref{fermi}) correctly reproduces the known result 
of the Fermi energy for conventional 3DEG:
$
E_F^0 = \frac{\hbar^2}{2m^\ast}(3\pi^2n_e)^{2/3}.
$
The topology of the Fermi surface changes at $n_e=n_t$ with $ n_t = 4k_{\alpha}^3/3\pi^2$.
For a given $\alpha$, $E_F < 0 $ when $ n_e < n_t $ 
and $E_F > 0 $ when $ n_e > n_t $. At finite temperature, 
we obtain the chemical potential $\mu$ by solving Eq. (\ref{Norm_C}) numerically 
for $\alpha=0.1$ eV-nm. For three different temperatures, namely, $T=5$, $10$, and 
$20$ K, the difference between the chemical potential and Fermi energy is shown 
in Fig. \ref{chemical}. We find that $\mu - E_F$ exhibits a dimensional crossover 
as the Fermi energy changes its sign. For instance, $\mu - E_F$ is negative when 
$E_F>0$. This feature corresponds to the nature of chemical potential in 3D case. 
On the contrary, for $E_F<0$, $\mu - E_F$ is positive. This behavior is clearly a hallmark of 
$\mu$ in 1D case. However, this indirect signature of dimensional crossover is not 
clearly seen from the structures of the density of states corresponding to 
$E_F>0$ and $E_F<0$.

\section{Thermoelectric transport}

\subsection{General formalism}
We consider the physical system is subjected to a spatially uniform electric 
field ${\bs\varXi}$ and a temperature gradient ${\bs \nabla}T$. The magnitude 
of the external electric field and temperature gradient are chosen in such a 
way that the linear response theory holds.

The electronic and thermal current densities are given by
${\bf j} = - e\sum_{\xi}{\bf v}_\xi f_\xi$
and 
${\bf j}_{\rm th} = \sum_{\xi}(E_\xi - \mu){\bf v}_\xi f_\xi$,
respectively. Here, $\xi$ defines the set of all quantum numbers, i.e., 
$\xi \equiv$ ($\lambda$, $\eta$, ${\bf k}$), and 
${\bf v}_\xi$ is the group velocity of the electron in the quantum state 
$\xi$.
Within the context of linear response theory, the  nonequilibrium distribution function 
$f_\xi$ is given by 
\begin{eqnarray}
\small{f_\xi=f_\xi^0+\tau_\xi \Bigg(-\frac{\partial f_\xi^0}{\partial E_\xi}\Bigg) 
{\bf v}_\xi \cdot
\Big[-e{\bs \varXi}^\ast+\frac{E_\xi-\mu}{k_BT}(-{\bs \nabla}T)\Big]},
\end{eqnarray}
where $f_\xi^0 = \Big[e^{(E_\xi-\mu)/(k_BT)} + 1 \Big]^{-1}$ is the Fermi-Dirac 
distribution function at equilibrium, $\tau_\xi \equiv \tau(E_\xi)$ is the energy 
dependent relaxation time,
and ${\bs \varXi}^\ast = {\bs \varXi} + {\bs \nabla}\mu/e$ is the effective electric field.

The current densities can be found together in the following form \cite{Ashcroft}:
\begin{eqnarray}
\begin{pmatrix}
{\bf j}\\
{\bf j}_{\rm th}
\end{pmatrix}
=
\begin{pmatrix}
L_{11} & L_{12}\\
L_{21} & L_{22}
\end{pmatrix}
\begin{pmatrix}
{{\bs \varXi}^\ast}\\
-{\bs \nabla}T
\end{pmatrix},
\end{eqnarray}
where $L_{pq}$'s are defined as 
$L_{11}=\mathcal{L}^{(0)}$, $L_{21}=TL_{12} = - \mathcal{L}^{(1)}/e$, 
and $L_{22}=\mathcal{L}^{(2)}/(e^2T)$ with 
\begin{eqnarray}\label{L_cal}
\mathcal{L}_{\nu\nu^{\prime}}^{(r)}=e^2\sum_\xi   \tau_\xi  v_{\nu}(\xi) v_{\nu^{\prime}}(\xi) (E_\xi-\mu)^r
\Bigg(-\frac{\partial f_\xi^0}{\partial E_\xi}\Bigg),
\end{eqnarray}
with $\nu,\nu^{\prime}=x,y,z$. Note that different combinations of $L_{pq}$ can be used to define the 
thermoelectric coefficients like electrical conductivity, thermopower, 
and thermal conductivity. For instance, $L_{11}$ can be identified as 
the electrical conductivity, whereas the thermopower and the thermal conductivity 
are defined as $ S = L_{21}/L_{11}$ and $\kappa = L_{22}-L_{21}(L_{11})^{-1}L_{12}$, 
respectively. We are going to determine $\mathcal{L}_{\nu\nu^{\prime}}^{(r)}$ in a more explicit way, 
relevant for the present scenario. Let us now restrict ourselves to 
consider that the scattering mechanisms, responsible for thermoelectric 
transport, are due to the presence of weak spin-independent disorders 
distributed throughout the sample with an average density
$n_{\rm imp}$. The short-range disorder potential is given by 
$U({\bf r})=U_0\sum_i\delta({\bf r}-\bf{r}_i)$, where $U_0$ is the 
strength of the potential having dimension of energy times volume and 
${\bf r}_i$ is the position of the $i$-th scatterer. Note that 
the present situation is an example of isotropic case since the energy spectrum 
depends only on the magnitude of ${\bf k}$. In this case, 
$\mathcal{L}_{\nu\nu^{\prime}}^{(r)}=\mathcal{L}_{\nu\nu}^{(r)}\delta_{\nu\nu^{\prime}}\equiv\mathcal{L}^{(r)}$. 
Also the expectation value of $\hat v_\nu({\bf k}) $ with respect 
to the state $\lambda$ is
$ \langle \hat{v}_{\nu}({\bf k})\rangle_{_ \lambda} = 
\frac{ \hbar k_{\nu}}{m^\ast} +\lambda \frac{\alpha}{\hbar} \frac{k_{\nu}}{k}$. 
Equation (\ref{L_cal}) can be rewritten further as
\begin{widetext}
\begin{eqnarray}\label{L_cal2}
\mathcal{L}^{(r)}&=&\frac{e^2}{(2\pi)^3}\sum_{\lambda=\pm} \int_{\mu>0}
d^3k \, \tau_{_\lambda}(E_{{\bf k}\lambda})\langle \hat{v}_{\nu}({\bf k})\rangle^2_{_{\lambda}}
(E_{{\bf k}\lambda}-\mu)^r 
\Bigg(-\frac{\partial f_{{\bf k}\lambda}^0}{\partial E_{{\bf k}\lambda}}\Bigg)\nonumber\\
&+&\frac{e^2}{(2\pi)^3}\sum_{\eta=1,2} \int_{\mu<0}
d^3k \, \tau_{_\eta}(E_{{\bf k}-})\langle \hat{v}_{\nu}({\bf k})\rangle^2_{_{-}}
(E_{{\bf k}-}-\mu)^r 
\Bigg(-\frac{\partial f_{{\bf k}-}^0}{\partial E_{{\bf k}-}}\Bigg).
\end{eqnarray}
\end{widetext}

The relaxation time $\tau_{_{\lambda(\eta)}}$
in Eq. (\ref{L_cal2}) can be determined using the 
framework of semiclassical Boltzmann transport theory including
interband scattering for multiband system.
When Fermi energy lies below the BTP, an unusual intraband scattering 
(interbranch and intrabranch scatterings) arises due to the concave-convex 
shaped Fermi surface topology as shown in Fig. \ref{sketch}(c).
For the present case, we solve the Boltzmann transport equation including 
the interband/interbranch scattering term self-consistently 
(see the Appendix for detail derivation) to find 
\begin{eqnarray}\label{tau1}
\tau_\lambda (E)=\frac{\hbar}{\pi n_{\rm imp}U_0^2}\Bigg[\frac{D_\lambda^>(E)}{[D_T^>(E)]^2}
+\frac{1}{2D_T^>(E)}\Bigg]
\end{eqnarray}
for $E>0$
and 
\begin{eqnarray}\label{tau2}
\tau_\eta (E)=\frac{\hbar}{\pi n_{\rm imp}U_0^2}\Bigg[\frac{D_\eta^<(E)}{[D_T^<(E)]^2}
+\frac{1}{2D_T^<(E)}\Bigg]
\end{eqnarray}
for $E<0$. Here, $D_T^>(E) = D_+^>(E) + D_-^>(E)$ and 
$D_T^<(E) = D_1^<(E) + D_2^<(E)$ are the total density of states for
$E>0$ and $-E_\alpha < E < 0$, respectively. 
Note that the expressions of $\tau(E)$
for $E>0$ and $E<0$ share the same mathematical structures. This help us to 
find $\mathcal{L}^{(r)}$ in a more compact form,
\begin{eqnarray}\label{NUM}
\mathcal{L}^{(r)}=\frac{4}{3}\frac{e^2}{hl_0}(k_BT)^r X_r,
\end{eqnarray}
where $l_0 = m^\ast n_{\rm imp}U_0^2/(\hbar^2k_BT)$,
$X_r=\int_{-x_\alpha}^\infty \mathcal{L}(x)(x - \tilde{\mu})^r dx$, with $x=E/(k_BT)$ and
$\tilde{\mu}=\mu/(k_BT)$. Here $\mathcal{L}(x)$ is given by
\begin{eqnarray}
\mathcal{L}(x)=f_x(1-f_x)(x+x_\alpha)\Bigg[1+2\frac{x_\alpha(x+x_\alpha)}{(x+2x_\alpha)^2}\Bigg],
\end{eqnarray}
with $f_x=[e^{x-\tilde{\mu}}+1]^{-1}$ and $x_\alpha = E_\alpha/(k_BT)$. 

\subsection{Results}
Here we discuss the behavior of different thermoelectric coefficients obtained via the 
numerical solution of Eq. (\ref{NUM}). For numerical calculation we consider following 
material parameters : effective mass of electron $m^\ast=0.5m_e$, $m_e$ is the free electron 
mass and $\alpha_0=10^{-10}$eVm. 

\begin{figure}[htbp]
\begin{center}\leavevmode
\includegraphics[width=0.49\textwidth]{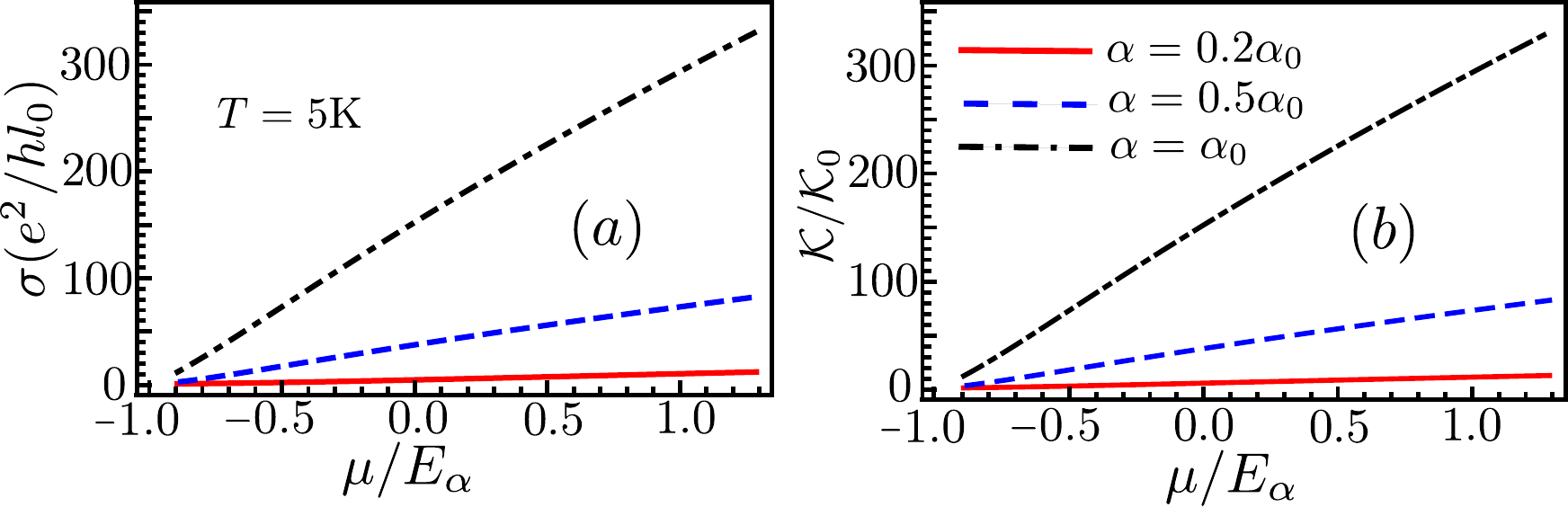}
\caption{Plots of (a) electrical conductivity $\sigma$ and (b) thermal conductivity 
$\kappa$ as a function of chemical potential $\mu$ at temperature $T=5$ K for 
three different values of $\alpha$.} 
\label{conductivities}
\end{center}
\end{figure}

Let us begin with the behavior of the electrical conductivity 
$\sigma = L_{11} = \mathcal{L}^{(0)}$.
From Eq. (\ref{NUM}), we explicitly have $\sigma=4e^2X_0/(3hl_0)$. 
Figure \ref{conductivities}(a) depicts the variation of 
$\sigma$ with chemical potential for $\alpha=0.2\alpha_0$, $0.5\alpha_0$, and 
$\alpha_0$ at $T=5$ K. The conductivity increases monotonically with $\mu$. 
For higher values of $\alpha$, the enhancement of $\sigma$ is significant. 
In the $T \rightarrow 0$ limit, we obtain the following analytical expression 
of $\sigma$ as (see the Appendix for detail derivation)
\begin{eqnarray}\label{Con_T0}
\sigma = \frac{e^2}{h} \frac{2 \hbar^2(E_\alpha + E_F)}{m^\ast n_{\rm imp} U_0^2}
\Big[ 1 - \frac{E_{F}^2}{3(2E_\alpha + E_F)^2} \Big],
\end{eqnarray}
which shows that the zero temperature conductivity increases linearly with the 
Fermi energy. Note that the characteristics of $\sigma$ with $E_F$ in the $E_F>0$ 
and $E_F<0$ are the same. This is different from the behavior of $\sigma$ with $E_F$ 
in 2D Rashba systems \cite{Uncon} in which it was found that the zero temperature 
conductivity depends on the Fermi energy in linear (quadratic) fashion for $E_F>0$ ($E_F<0$). 
This feature was attributed to the fact that $\sigma(E)$ cannot be continuously 
differentiated at $E_F=0$. In that 2D Rashba system density of states of individual 
bands/branches depends linearly on wave vector, which immediately implies that total 
density of states below and above BTP have different energy dependence. So relaxation 
time having similar structures below and above the BTP will have different energy dependence.  
As a result of this the electrical conductivity being proportional to the relaxation time 
shows different energy dependence below and above BTP. However, in the present context, Eq. (\ref{Con_T0}) 
clearly demonstrates that $\sigma(E)$ is a continuously differentiable function at 
$E_F=0$. This fact helped us to find same analytical structures of $\sigma(E_F)$ in 
both $E_F>0$ and $E_F<0$ regions. 
From Eq. (\ref{Con_T0}) it is evident that 
$\sigma \propto E_\alpha$ at $E_F=0$. 
The signature of this feature is also reflected 
in Fig. \ref{conductivities}(a).
The reason of not seeing the direct indication of change in Fermi surface topology
in electrical conductivity in our case is that the density of states
depends on square of the wave vector as a consequence of the  3D system.
It implies that the form of total density of states, and hence relaxation time will have the 
same energy dependence below and above
the BTP (see the Appendix for more details).

The thermal conductivity $\kappa$ is obtained as $\kappa=(4/\pi^2)\kappa_0\kappa_I$,
where $\kappa_0=e^2 L_0T/(hl_0)$, with $L_0=(\pi^2/3)(k_B/e)^2$ as 
the Lorentz number for $3$D electron gas and $\kappa_I=X_2-X_1^2/X_0$. The variation of 
the thermal conductivity with the chemical potential is depicted in 
Fig. \ref{conductivities}(b). 
The thermal conductivity behaves with the chemical potential in a similar fashion as 
the electrical conductivity.

\begin{figure}[htbp]
\begin{center}\leavevmode
\includegraphics[width=0.45\textwidth]{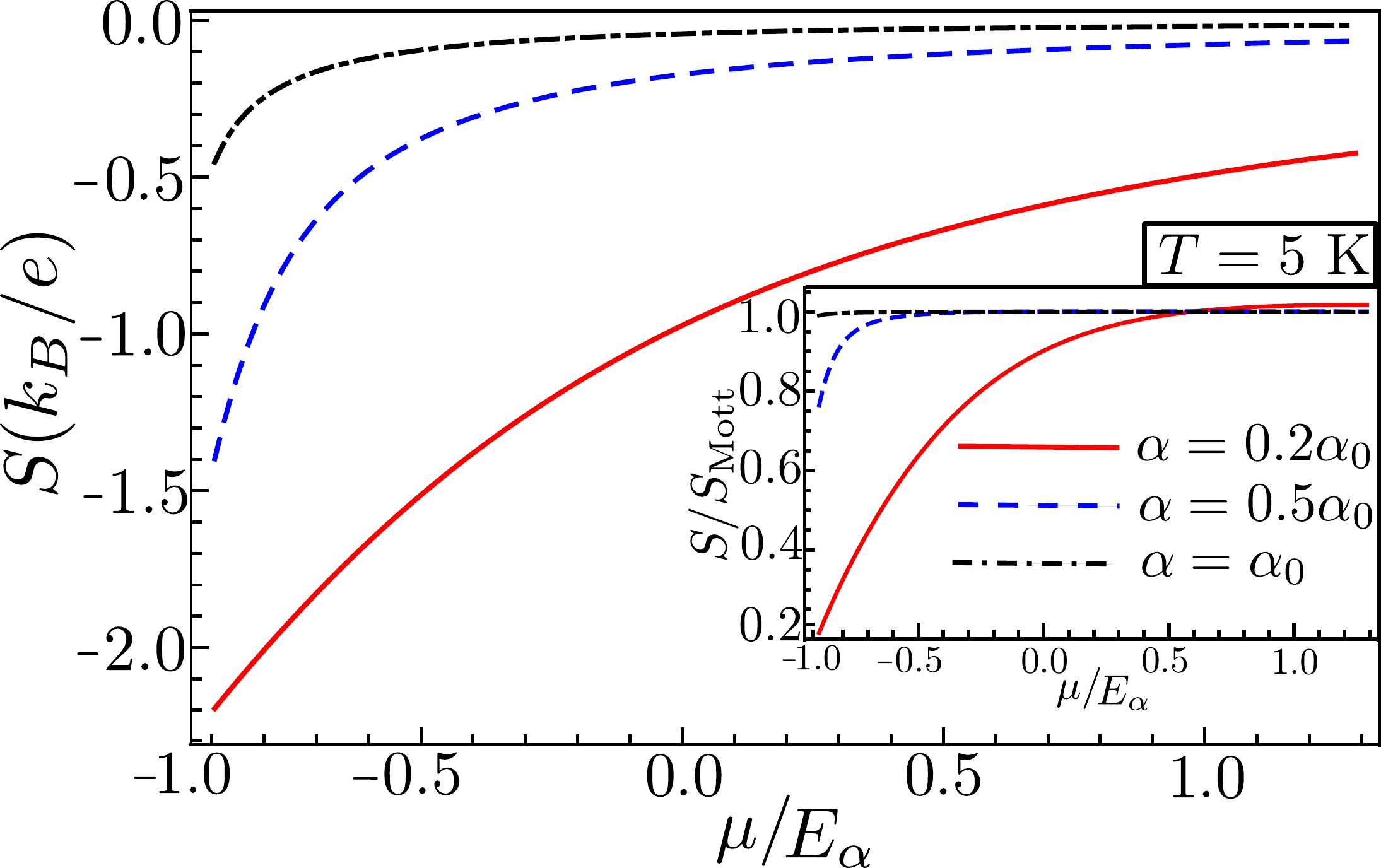}
\caption{Plots of the thermopower $S$ as a function 
of chemical potential $\mu$ at temperature $T=5$ K for three different values of $\alpha$.}
\label{thermopower}
\end{center}
\end{figure}

We obtain the thermopower through explicit calculation as
$S = - (k_B/e)S_I$, where $S_I=X_1/X_0$. In Fig. \ref{thermopower}, we show the 
dependence of $S$ on $\mu$
for $\alpha=0.2\alpha_0$, $0.5\alpha_0$, and $\alpha_0$. The thermopower is large at lower
values of $\alpha$ as compared to higher $\alpha$. In the region below $E_F=0$ magnitude of thermopower 
changes rapidly with $\mu$. For $E_F>0$ the rate of change of $S$ with $\mu$ is slow 
compared to the previous case. Moreover, for higher $\alpha$, the thermopower attains a 
saturation value when $\mu>0.5E_\alpha$. In the 
$\vert E_F\vert\gg k_BT$ limit, using the Mott relation \cite{Ashcroft}
\begin{eqnarray} \label{Mott-relation}
S_{\rm Mott} = - \frac{\pi^2 k_B^2T}{3e}\frac{d{\rm ln}\sigma(E)}{dE}\Big\vert_{E=E_F},
\end{eqnarray}
one may obtain the following expression of thermoelectric power:
\begin{eqnarray} \label{Mott}
S_{\rm Mott}=-\frac{\pi^2 k_B T}{3e(E_F+E_\alpha)}\Bigg[1-
\frac{2E_F E_\alpha(E_F+E_\alpha)}{(E_F+2E_\alpha)\xi_F}\Bigg],
\end{eqnarray}
where $\xi_F=E_F^2+6E_\alpha E_F+6E_\alpha^2$. The inset of 
Fig. \ref{thermopower} shows the variation of
$S/S_{\rm Mott}$ as a function of chemical potential for 
different values of $\alpha$ at $T=5$ K. For lower values of $\alpha$, namely, 
$\alpha=0.2\alpha_0$, the Mott relation is valid only when $\mu > 0.5E_\alpha$. 
The degree of validity of the Mott relation increases with $\alpha$. For $\alpha=\alpha_0$, 
the Mott formula is satisfied almost in the entire range of the chemical potential considered. 
It is known \cite{Ashcroft,Uncon} that the validity of Mott relation or the Sommerfeld 
expansion depends on the following simultaneous conditions: (1) $T\ll T_F$ where 
$T_F=E_F/k_B$ is the Fermi temperature and (2) whether the Taylor expansion of 
$\sigma(E)$ about $E=\mu$ is possible or not. In our case, Condition (2) is satisfied 
always because $\sigma(E)$ is a continuously differentiable function. Therefore, we 
attribute the breakdown of Mott relation (as shown in Fig. \ref{thermopower}) to 
the breakdown of the validity of Condition (1).

\begin{figure}[htbp]
\begin{center}\leavevmode
\includegraphics[width=0.45\textwidth]{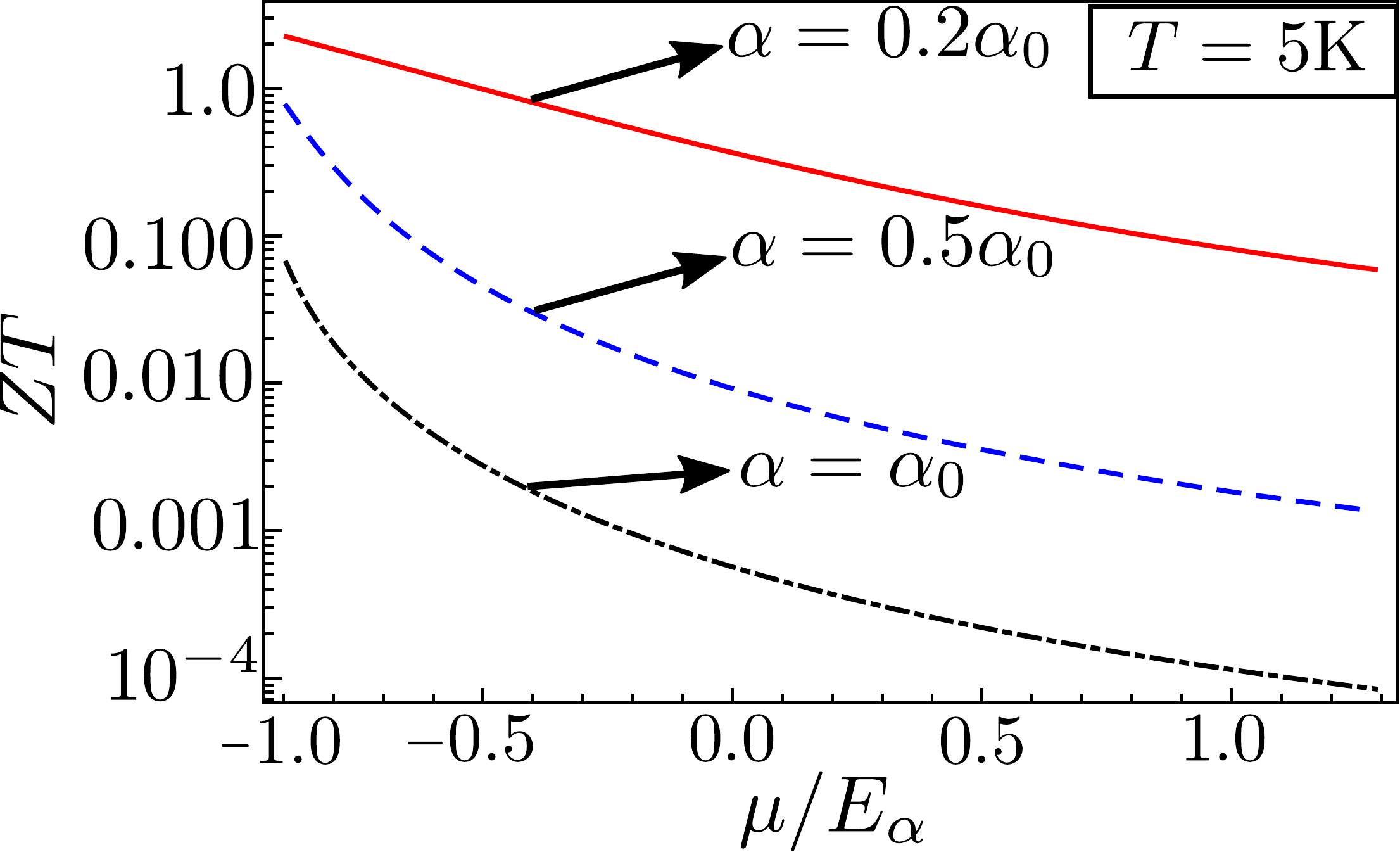}
\caption{Plots of the thermoelectric figure of merit $ZT$ as a function 
of chemical potential $\mu$ at temperature $T=5$ K for three different values 
of $\alpha$.}
\label{fom}
\end{center}
\end{figure}

The thermoelectric figure of merit is a dimensionless number which measures the 
efficiency of the thermoelectric performance of a material. It is defined as
$ZT=S^2\sigma T/\kappa$. In general, the symbol $\kappa$ used in the 
definition of $ZT$ stands for the total thermal conductivity which is a sum of 
electronic and phononic contribution. But, in the temperature regime we are focusing on now, 
the electronic thermal conductivity dominates over the thermal counterparts. Therefore, 
we neglect lattice contribution to the thermal conductivity and proceed with the electronic 
counterpart for the subsequent calculation of $ZT$.
A good thermoelectric material needs to possess following 
properties : large electrical conductivity, high thermopower, and low thermal 
conductivity. 
For the current scenario, we obtain $ZT$  explicitly as 
$ZT = S_I^2X_0/\kappa_I$. In Fig. \ref{fom} we show the dependence of $ZT$ on the 
chemical potential at $T=5$ K for three different values of $\alpha$, namely, 
$\alpha=0.2\alpha_0$, $0.5\alpha_0$, and $\alpha_0$. It is clear that $ZT$ behaves 
as a monotonically decreasing function of $\mu$ for each $\alpha$. The value of $ZT$ 
is higher at lower $\alpha$. This is obvious because the order of magnitude of 
$\sigma$ and $\kappa$ are almost same for each $\alpha$ whereas the thermopower 
is large at lower $\alpha$. In other words, the thermopower alone determines the 
behavior of $ZT$.

\begin{figure}[htbp]
\begin{center}\leavevmode
\includegraphics[width=0.45\textwidth]{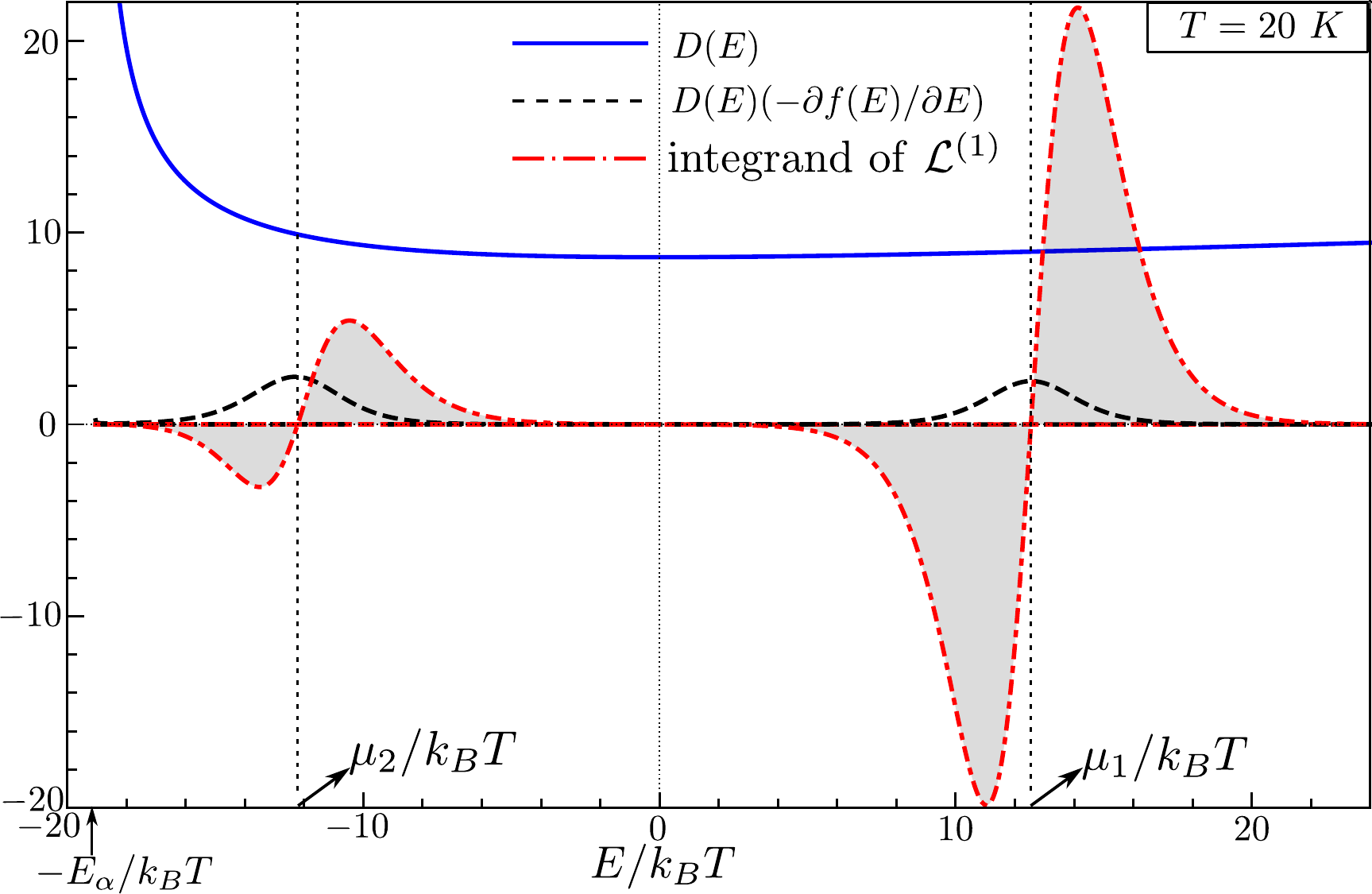}
\caption{Plots of $D(E)$, $D(E)(-\partial f^0/\partial E)$, and the integrand 
of $\mathcal{L}^{(1)}$ as a function of $E$ for $\alpha=\alpha_0$. For a better 
visualization we consider $T=20$ K.}
\label{explain}
\end{center}
\end{figure}

Note that for $\alpha=0.2\alpha_0$, $ZT$ attains a remarkable value 
greater than $2$ when the chemical potential lies far below the BTP. 
To explain this feature explicitly, in Fig. \ref{explain} we plot 
$D(E)$, $D(E)(-\partial f^0/\partial E)$, and the integrand of $\mathcal{L}^{(1)}$ 
(which is proportional to $S$) as a function of energy for a particular $\alpha$, 
namely, $\alpha=\alpha_0$. We consider two different 
densities, namely, $n_{e}^{(1)}$ and $n_{e}^{(2)}$ which in turn correspond to chemical 
potential $\mu_1$ and $\mu_2$, respectively, at some constant temperature $T$. 
The chemical potential $\mu_1$ lies above the band touching point whereas $\mu_2$
falls well below of it. As expected the function $D(E)(-\partial f^0/\partial E)$ exhibit 
a peak whenever the energy 
matches with the chemical potential.
For $n_e =n_{e}^{(1)}$, as expected the integrand of 
$\mathcal{L}^{(1)}$ changes 
sign when the energy crosses $\mu_1$ and exhibits a structure as shown by the shaded 
portion. It is hard to differentiate between the areas under the curves (shaded region in Fig. \ref{explain}) below 
and above $\mu_1$. Therefore, when summed up it gives rise to negligible contribution 
to the thermoelectric power which is reflected in Fig. \ref{thermopower}.

When $n_e =n_{e}^{(2)}$, the asymmetry between the 
magnitudes of $\mathcal{L}^{(1)}$  below and above $\mu_2$ can be visible. This increment in the asymmetry is responsible 
for the enhancement of thermopower below BTP. The amount of asymmetry increases as the chemical potential
approaches $-E_\alpha$. As we tune the chemical potential from positive to negative value the amount 
of asymmetry increases due to the dimensional crossover from 3D to 1D, which leads to an increase of 
thermopower as well as thermoelectric figure of merit. There have been several studies 
before\cite{MSD,MSD1}, where it has been shown that the low dimensional systems lead to larger 
asymmetry in the density of states about the chemical potential, which give rise to larger thermopower 
and thermoelectric figure of merit and hence can be used for good thermoelectric device applications. 
Our result is consistent with these observations as we obtain larger value of figure of merit below 
the BTP due to the underlying 1D-like characteristics of our bulk 3D system. It can be verified that 
the amount of asymmetry will become more 
prominent for the cases of lower $\alpha$. This asymmetry explains the higher values 
of ZT obtained for low $\alpha$. Therefore, one has to struggle to fix both $\alpha$ 
and $n_e$ at
reasonable values  to use 3D noncentrosymmetric metal for good thermoelectric devices. 
Also the  thermopower $S$ and figure of merit $ZT$
tend to diverge as $\mu$ approaches $-E_\alpha$. This fact is
attributed to the presence of the van Hove singularity in the band structure,
similar to quasi-1D systems \cite{MSD1}.

We emphasize here that at very low density (as $E_F \rightarrow - E_\alpha$), 
	the singularity in the density of states may be broadened by disorder due 
	to finite-band effects \cite{quantumeffects2DEG}.
	Such quantum effects must be considered using the Kubo formula \cite{mahan}, rather than
	semiclassical Boltzmann transport theory.

\section{Zero-momentum optical conductivity}
\begin{figure}[htbp]
\begin{center}\leavevmode
\includegraphics[width=0.48\textwidth]{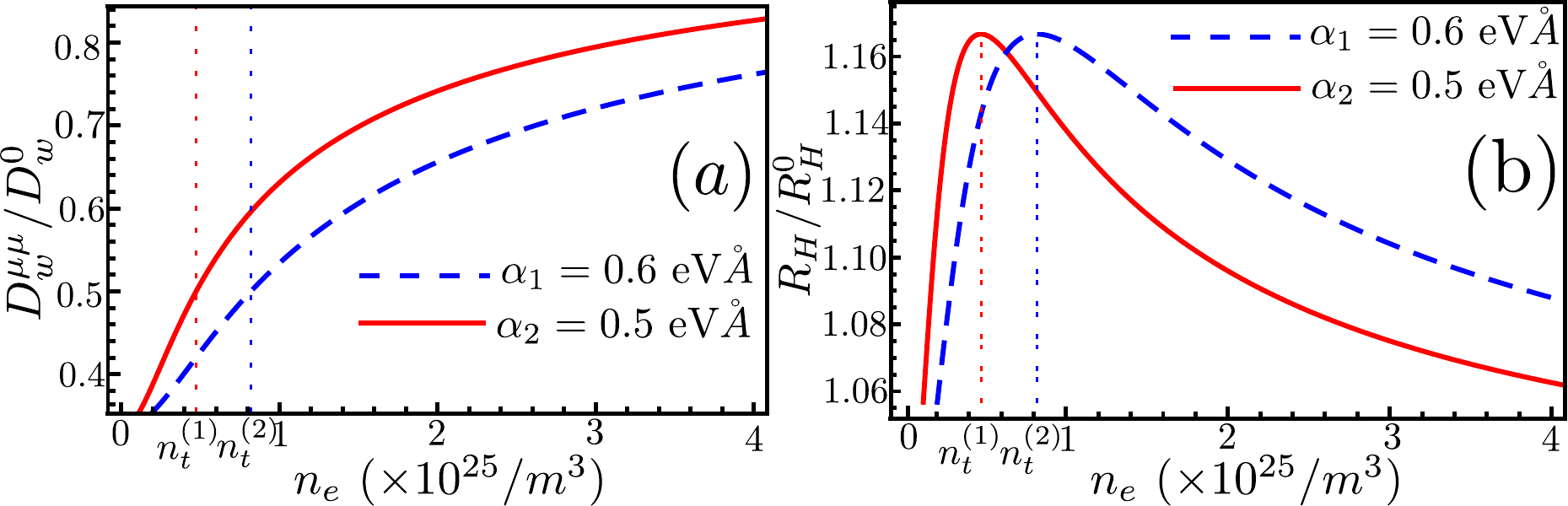}
\caption{ Plots of the Drude weight $D_{w}^{\nu \nu^{\prime} }$ (a) and the Hall coefficient 
$R_{H}$ (b) as a function of carrier density $n_e$ for two different values of $\alpha$. 
Here $n_{t}^{(i=1,2)} = 4 k_{\alpha_{i}}/(3\pi^2)$ (with $i=1,2$) is the density at 
which $E_{F}$ changes its sign, i.e., the topology of the Fermi surface changes at 
$n_{t}$.}
\label{Drude}
\end{center}
\end{figure}

In this section, we present optical signature of a change in the
Fermi surface topology in noncentrosymmetric metals at $T=0$.
The zero-frequency and finite-frequency optical conductivities are 
the manifestations of the intra-band and interband optical transitions, respectively. 
Consider the noncentrosymmetric metal is irradiated by a weak and 
spatially homogeneous electric field 
${\bf E} = E_0 e^{i\omega t} \hat {\bs \nu}$ 
oscillating with the frequency $\omega$ and amplitude $E_0$. 
The absorptive part of the charge optical conductivity tensor is given by 
${\rm Re} \; \Sigma_{\nu \nu^{\prime} }(\omega) = \delta(\omega) D_w^{\nu \nu^{\prime}}
+  {\rm Re} \; \sigma_{\nu \nu^{\prime} }(\omega)$,
where $\nu, \nu^{\prime}=x,y,z$ and 
$D_w^{\nu \nu^{\prime} }$ is the Drude weight (charge stiffness).
The semiclassical expression of the Drude weight \cite{Ashcroft} is given by
\begin{eqnarray} \label{drude-gen}
D_{w}^{\nu \nu^{\prime} } = \pi e^2\sum_{\lambda}\int \frac{d^3 k}{(2\pi)^3}
\langle\hat{v}_{\nu}\rangle_{\lambda}\langle \hat{v}_{\nu^{\prime} }\rangle_{\lambda} 
\delta(E_\lambda({\bf k}) - E_F).
\end{eqnarray}
This semiclassical expression has been successfully used in 2DEG with linear 
spin-orbit interaction as well as in 2D hole gas with $k$-cubic 
spin-orbit interaction \cite{Alestin}.
The exact analytical expression of the Drude weight is obtained as
\begin{eqnarray}\label{drude}
D_{w}^{\nu \nu^{\prime} } & = & 
\frac{\pi e^2 n_e }{m^\ast} \Big[\frac{E_F + 2 E_{\alpha} }{E_F + 4 E_\alpha} \Big] 
\delta_{\nu \nu^{\prime} } \theta(E_F + E_\alpha). 
\end{eqnarray}
The longitudinal Drude weight is isotropic, i.e., 
$ D_{w}^{xx} = D_{w}^{yy} = D_{w}^{zz} = D_w$ and 
the off-diagonal Drude weight vanishes exactly. 
Setting $\alpha=0$ in the above equation, we get the standard result of 
the Drude weight for 3DEG, i.e., $D_{w}^{0} = \pi n_e e^2/m^\ast$.
It shows that the spin-orbit coupling reduces the Drude weight as compared to the 
conventional 3DEG without spin-orbit coupling and further an increase in $\alpha$ 
decreases it more as shown in Fig. \ref{Drude}(a). We also see in Fig. \ref{Drude}(a) that 
similar to Drude conductivity, 
Drude weight also shows no signature of change in Fermi surface topology.

It has been shown \cite{zotos,zotos1} that the Hall coefficient $R_H$ can be 
obtained from the Drude weight using the general expression
$$
R_H = - \frac{1}{e D_w} \frac{\partial D_w}{\partial n_e}.
$$
This expression has been successfully used to calculate the Hall
coefficient in various systems \cite{zotos,zotos1,hall-graphene}.
For the present system, the exact Hall coefficient is given by
\be \label{hall}
R_H =  R_{H}^0
\Big[1 + \frac{2 E_{\alpha} (E_{\alpha} +E_F)}
{3(2E_{\alpha}+ E_F)^2} \Big]  \theta(E_F + E_\alpha),
\ee
where $ R_{H}^0 = - 1/(e n_e)$ is the Hall coefficient 
for $\alpha =0 $ case. Equation (\ref{hall}) clearly shows the Hall 
coefficient is enhanced due to the presence of the spin-orbit coupling. 
In the low-density limit, 
$E_F$ is comparable to $E_\alpha$ and 
therefore pronounced effect of the spin-orbit coupling can be realized when 
$E_F < 0$ as seen in Fig. \ref{Drude}(b). It is interesting to notice here that Hall 
coefficient responds to the change in the Fermi surface topology. As shown in 
Fig. \ref{Drude}(b), it first rises sharply until $n_{t}$ where transition 
occurs and then starts decreasing with further increase  in density.
The peak in $R_H/R_H^0$ at the BTP is the signature of the change in the Fermi
surface topology. Generally we estimate carrier concentration from the Hall coefficient 
measurement. It is interesting to note that for noncentrosymmetric metals we can also estimate 
the strength of RSOI from this measurement by noting the transition density $n_t$ which depends on $\alpha$.

The finite-frequency optical conductivity $\sigma_{\nu \nu^\prime}(\omega)$ is 
arising due to the transitions between the spin-split states.
Within the linear response Kubo formalism,  
the frequency-dependent optical conductivity is given by
\begin{eqnarray}
\sigma_{\nu \nu^{\prime} }(\omega) & = & \frac{1}{\hbar(\omega + i 0^+)} 
\int_{0}^{\infty}dt  e^{i(\omega + i 0^+)t} 
\langle [\hat j_{\nu}(t), \hat j_{\nu^{\prime} }(0) ]  \rangle \nn.
\end{eqnarray}
Here $\hat j_{\nu} = e \hat v_{\nu} $ are the components of the 
charge current density operator,
$$
\langle [\hat j_{\nu}(t), \hat j_{\nu^{\prime} }(0) ] \rangle = 
\sum_{\lambda} \int d^3k \la \lambda {\bf k}| [\hat j_{\nu}(t), \hat j_{\nu^{\prime}}(0) ]| 
\lambda {\bf k} \rangle
f(E_{{\bf k}\lambda}) 
$$ 
denotes the quantum and thermal average in the interaction picture,
$\la {\bf r}| \lambda {\bf k} \ra = \psi_{\bf k}^{\lambda}({\bf r})$ 
and $f(E_{{\bf k}\lambda})$ is the Fermi-Dirac distribution function.

\begin{figure}[htbp]
\begin{center}\leavevmode
\includegraphics[width=0.48\textwidth]{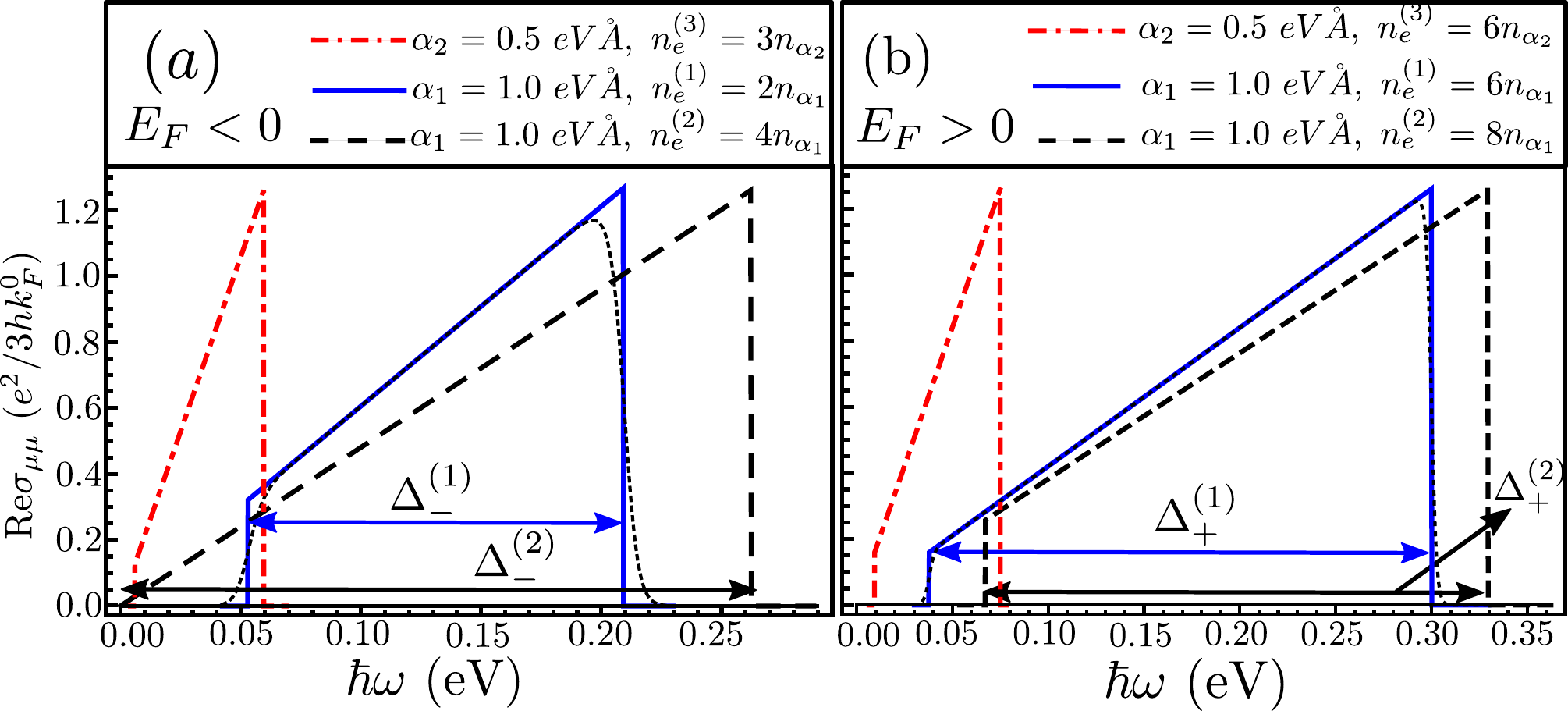}
\caption{(a) Plots of the Re $\sigma_{\nu \nu}$ as a function 
of $\hbar\omega$ for $E_{F}<0$. Solid and dashed curves corresponds 
to same $\alpha$ but different density and have optical absorption widths 
$\Delta_{1}^-$ and $\Delta_{2}^-$, respectively. Here $\Delta_{-}^{(1)} < \Delta_{-}^{(2)}$ 
as $n_{e}^{(1)} < n_{e}^{(2)} $, so $\Delta$ depends on density as mentioned in the text. 
Dotted-dashed curve corresponds to different values of $\alpha$ and $n_e$, 
here $\Delta$ becomes more narrow because 
of the smaller value of $\alpha$. 
(b) Plots of the Re $\sigma_{\nu \nu}$ as a 
function of $\hbar\omega$ for $E_{F}>0$. Solid and dashed curves corresponds 
to same $\alpha$ but different densities and have optical absorption widths 
$\Delta_{+}^{(1)}$ and $\Delta_{+}^{(2)}$, respectively. 
Here $\Delta_{+}^{(1)} = \Delta_{+}^{(2)}$  but $n_{e}^{(1)} < n_{e}^{(2)}$, 
so $\Delta$ is independent of carrier density as mentioned 
in the text. Dotted-dashed curve corresponds to different values of $\alpha$ and $n_e$, 
here $\Delta$ becomes more narrow because of the smaller value of $\alpha$. The dotted 
thin line in both the parts shows the finite temperature behavior of the optical conductivity 
for $\alpha_1$ and $n_{e}^{(1)}$. }
\label{optical}
\end{center}
\end{figure}
After some straight forward calculation, 
the real part of charge optical conductivity is given by
\begin{eqnarray}\label{opt}
{\rm Re \; \sigma_{\nu \nu^{\prime} }(\omega)} & = & \frac{e^2}{8\pi^2\omega} 
\int d^3k[f(E_{{\bf k}-})-f(E_{{\bf k}+})] v_{\nu}^{-+}({\bf k}) \nonumber \\
& \times & v_{\nu^{\prime} }^{+-}({\bf k}) 
\delta(E_{{\bf k}+}-E_{{\bf k}-}-\hbar \omega),
\end{eqnarray}
where
\begin{eqnarray}
v_{x}^{+-}({\bf k}) & = & 
\frac{\alpha}{\hbar}[e^{-2i\phi}\sin^{2}(\theta/2)-\cos^{2}(\theta/2)],\\
v_{y}^{+-}({\bf k})&=&\frac{ i\alpha}{\hbar}[e^{-2i\phi}\sin^{2}(\theta/2) 
+ \cos^{2}(\theta/2)],\\
v_{z}^{+-}({\bf k})&=&\frac{\alpha}{\hbar}e^{-i\phi}\sin\theta,
\end{eqnarray}
and $ v_{\nu}^{-+}({\bf k}) = [v_{\nu}^{+-}({\bf k})]^*$. 
The only root of the equation 
$E_{{\bf k}+}-E_{{\bf k}-} - \hbar \omega=0$ is 
$k_{\omega}=\hbar\omega/2\alpha$. 
Using the result of the following angular integrations,
\begin{eqnarray}
\int_{0}^{\pi}\int_{0}^{2\pi} v_{\nu}^{+-}({\bf k})v_{\nu^{\prime}}^{-+}({\bf k}) 
\sin\theta d\theta d\phi 
= \frac{8 \pi \alpha^2}{3 \hbar^2} \delta_{\nu \nu^{\prime} },
\end{eqnarray}
the final expression of real part of the optical conductivity at $T=0$ is given by,
\begin{eqnarray} \label{op-con-final}
{\rm Re \; \sigma_{\nu \nu^{\prime} }(\omega)}
& = & \frac{e^2}{h} \frac{k_{\omega}}{3}
\Big[\theta(2\alpha k_{F}^{-}-\hbar\omega) \theta(\hbar\omega-2\alpha k_{F}^{+}) 
\theta(E_F) \nonumber \\  
& + &  \theta(2\alpha k_{2}-\hbar\omega) \theta(\hbar\omega-2\alpha k_{1}) \nonumber \\
& \times & \theta(-E_F) \theta(E_F + E_\alpha) \Big] \delta_{\nu \nu^{\prime} }.
\end{eqnarray}

Equation (\ref{op-con-final}) shows the isotropic nature of the longitudinal optical
conductivities:
$ {\rm Re \; \sigma_{xx}(\omega)} = {\rm Re\;\sigma_{yy}(\omega)} 
= {\rm Re\;\sigma_{zz}(\omega)} \equiv  {\rm Re\;\sigma (\omega)} $ and
absence of the off-diagonal conductivities: 
$ {\rm Re\;\sigma_{\nu \nu^{\prime} }(\omega)} = 0 $ for $\nu \neq \nu^{\prime} $.

For $E_F \geq 0$, the interband optical transitions would occur when the photon 
energy ($\hbar \omega$) obeys the inequality 
$ 2\alpha k_{+}^F \leq \hbar \omega \leq  2\alpha k_{-}^F$ at $T=0$. 
However, for $E_F \leq 0$, the interband optical transitions take place
when the photon energy satisfies the inequality 
$ 2\alpha k_{1}^F \leq \hbar \omega \leq  2\alpha k_{2}^F$ at $T=0$.
The optical absorption width, the region where ${\rm Re} \; \sigma(\omega)$ remains nonzero, 
is $\Delta_+ = 8 E_{\alpha} $ for $E_F \geq 0$ and 
$\Delta_- = 8 \sqrt{E_{\alpha}^2 + E_{\alpha} E_F}$ for 
$E_F \leq 0 $. Interesting to note that 
$\Delta_+ $ is independent of the carrier density, whereas 
$ \Delta_-$ depends on the both carrier density as well as Rashba 
energy $E_\alpha$ as shown in Fig. \ref{optical}. 
The density dependence of $ \Delta_-$ is attributed to the topological 
change in the Fermi surface. So the optical conductivity shows a distinct response 
to the change in the Fermi surface topology. The absorption widths can be used to 
determine the value of $\alpha$ experimentally.
The magnitude of the optical conductivity at the left and right edges, respectively,
are
$ \sigma_{L/R}^+  =  \frac{e^2 k_{-/+}^{F}}{3\hbar} $ for $ E_F \geq 0$ and
$ \sigma_{L/R}^-  = \frac{e^2 k_{2/1}^{F}}{3\hbar} $ for $ E_F \leq 0$. 
We observe that $ \sigma_L^+ <  \sigma_R^+$ and $ \sigma_L^- <  \sigma_R^-$ which 
is also clear from Fig. \ref{optical}.

It is interesting to note that the spin-split energy gap is of the order 
of 0.1 eV for the carrier density $n_e = 10^{25}$ m${}^{-3}$ and 
$\alpha= 1.0 \times 10^{-10} $ eV-m. This energy scale is 
comparable to the electromagnetic radiation with 
very high frequency $\omega \sim 10^{14}$ Hz.
The high-frequency radiation would flip the spin
in a very short time. The noncentrosymmetric semiconductors 
can be used for high-speed spintronic devices. 
It should be emphasized here that the spin-orbit coupling locks electron's spin with its momentum 
	which changes due to scattering from impurities. Hence the charge 
	carrier's spin can flip due to strong spin-orbit coupling.
	Typically, the spin scattering rate for Elliot-Yafet and Dykonov-Perev
	mechanisms are \cite{Zutic}
	$\tau/\tau_{s}^{EY} \sim (\Delta_{\rm so}/{E_F})^2$ and 
	$1/\tau_{s}^{DP} \sim  \tau E_{\alpha} E_F /\hbar^2$, respectively.
	Here $\Delta_{\rm so} = \alpha k_F$ and $\tau$ being the momentum scattering time.
	Therefore, the spin-flip scattering rate
	can be reduced for suitable choice of moderate density and weak spin-orbit coupling,
	which will be the criteria for this system to use it for good spintronic device applications
at low temperature.


\section{Summary and conclusions}
In summary, we have theoretically studied signatures of the Fermi surface 
topology change in thermoelectric and optical properties of noncentrosymmetric 
metals.
The noncentrosymmetric metals possess distinct Fermi surface topology which depends
on the sign of the Fermi energy. As a result of this, the chemical
potential is found to exhibit a dimensional crossover from 3D to 1D-like
behavior as the Fermi energy switches its sign from positive to negative
one. It is shown that the electrical conductivity is continuously differentiable 
at the band touching point, as opposed to 2D Rashba systems. 
There is a significant enhancement of thermopower in the low density regime 
which is responsible to obtain a remarkable thermoelectric figure of merit with
value more than 2. However the figure of merit is found to decrease with
the increase of the strength of the Rashba spin-orbit interaction. This feature
is explained qualitatively. It is shown that the Hall coefficient
first rises sharply until BTP and then starts decreasing with further increase 
in density. 
The absorption width above the BTP depends solely on the spin-orbit 
coupling strength. Hence Hall coeffiecient and optical conductivity measurements 
can be used to extract $\alpha$ experimentally.
However, the absorption width below the BTP depends on both the density and
$\alpha$. 
The spin-split energy gap is comparable to the electromagnetic radiation with 
high frequency $\omega \sim 10-100$ THz. The corresponding spin-flip time 
scale will be very small.
Therefore, noncentrosymmetric bulk materials can be used for good thermoelectric
as well as spintronics devices with appropriate system parameters at low temperatures.

\section{Acknowledgement}
T.B. sincerely acknowledges the financial support provided 
by the University of North Bengal to pursue this work. 
We also acknowledge Dr. Arijit Kundu for some useful discussions.

\appendix{}
\begin{widetext}
\section{}
\subsection{Calculation of the relaxation time}
{\bf Relaxation time approximation for multiband systems}: In this Appendix, 
we present calculation of the relaxation time by solving the Boltzmann transport 
equation including the interband scattering for $ E>0$ and interbranch scattering 
for $E<0$ self-consistently.
We consider electrons in noncentrosymmetric semiconductors with spin-independent
short-range scatterer. This system is subjected to a spatially uniform electric field
${\bs \varXi}$ and temperature gradient ${\bs \nabla} T$.
The effective electric field due to charge redistribution results from ${\bs \varXi} $ is
given by
${\bs \varXi}_{\rm eff} = {\bs \varXi} - ({\bs \nabla} \mu)/e$. We now
linearize the Fermi-Dirac distribution function 
$f_\xi(E_\xi) =  f_{\xi}^0(E_\xi) + \delta f_\xi $
around the equilibrium solution $ f_{\xi}^0(E_\xi)$. Here 
$\xi \equiv (\lambda, {\bf k} )$ for $E \geq 0 $ and $\xi \equiv (\eta, {\bf k} )$ 
for $-E_{\alpha} \leq E \leq 0 $ is the eigenstate index; and 
$ \delta f_\xi $ is the  out-of-equilibrium 
deviation which is linear in the external electric field.
In nonequilibrium steady states, the linearized Boltzmann transport equation \cite{Ziman}
for the charge carriers is
\begin{eqnarray} \label{bte}
{\bf F}_\xi \cdot {\bf v}_\xi \frac{\partial f^0}{\partial E_\xi} 
= - \sum_{\xi^{\prime}} W_{\xi^{\prime},\xi}(\delta f_\xi - \delta f_{\xi^{\prime}}).
\end{eqnarray}
Here the generalized force acting on the state $\xi$ is
\begin{eqnarray}
{\bf F}_\xi = - \frac{(E_\xi - \mu)}{T} {\bs \nabla} T + e {\bs \varXi}_{\rm eff},
\end{eqnarray}
and ${\bf v}_\xi$ is the group velocity of the state $\xi$.  
Also, $ W_{\xi^{\prime},\xi} $ is the transition rate from the state $\xi$
to the state $\xi^{\prime}$. We have used the fact that
$   W_{\xi^{\prime},\xi}  = W_{\xi,\xi^{\prime}}$ in the Boltzmann transport equation.
Within the lowest-order Born approximation,
the intra-band transition rate between the states $\xi$ and $\xi^{\prime}$ is
$$
W_{\xi^{\prime} \xi} = \frac{2\pi N_{\rm imp} }{\hbar}
\big| \big\langle  \phi_{\xi^{\prime}}({\bf k}^{\prime})
\big|U({\bf r})  \big| \phi_{\xi}({\bf k}) \rangle\big|^2
\delta(E_{\xi} - E_{\xi^{\prime}} ).
$$
Here, $N_{\rm imp}$ is the number of $\delta$-scatterer randomly 
distributed in the system at various locations ${\bf r}_i$. The corresponding
spin-independent impurity potential produced by the
$\delta$-scatterer is given by 
$ U({\bf r}) = U_0 \sum_{i=1}^{N_{\rm imp}} \delta({\bf r} - {\bf r}_i)$.
Here $U_0$ being the strength of the impurity potential, whose dimension is energy
times volume. The Fourier transform of the potential $ U({\bf r})$ is
$ U({\bf q}) = U_0$.
Upon simplification, the transition rates are obtained as
\begin{eqnarray}
W_{\lambda^{\prime} \lambda} & = & \frac{1}{u_0 V} (1+ \lambda \lambda^{\prime} 
\cos \theta) \delta(E_\lambda - E_{\lambda^{\prime}}); \hspace{0.64cm} E \geq 0 ,\\
W_{\eta^{\prime} \eta} & = & \frac{1}{u_0 V} (1+ \cos \theta)
\delta(E_\eta - E_{\eta^{\prime}}); \hspace{0.4cm} - E_{\alpha} \leq E \leq 0 ,  
\end{eqnarray}
where $ 1/u_0 = \pi n_{\rm imp} U_{0}^2/\hbar $ with 
$n_{\rm imp}$ being the impurity density.


We need to solve the Boltzmann transport equation separately 
for $E \geq 0$ and $-E_{\alpha} \leq  E \leq 0$. This is because the 
intraband and interband transitions  take place when $E \geq 0$. Whereas intrabranch 
and interbranch transition occurs within the band $\lambda = - $.
We are able to obtain exact analytical expression of the scattering time even if 
we keep the interband/interbranch contribution in the Boltzmann transport equation.  
Assuming the out-of-equilibrium distribution function is of the following form:
\begin{eqnarray} \label{ansatz}
\delta f_{\xi}(E, {\bf v}(E_\xi,\theta,\phi)) = - \frac{\partial f^0(E)}{\partial E}
{\bf F}_\xi \cdot {\bf v}_{\xi}(E, \theta,\phi) \tau_\xi(E).
\end{eqnarray}
	
Substituting Eq. (\ref{ansatz}) into Eq. (\ref{bte}), the self-consistent equation 
for the relaxation time $\tau_{\xi}(E)$ is
\begin{eqnarray}
\frac{1}{\tau_{\xi}(E)} & = & \frac{V}{4 \pi} \sum_{\xi^{\prime}} 
\int dE_{\xi^{\prime}} D_{\xi^{\prime}}(E_{\xi^{\prime}})
\sin\theta^{\prime} d\theta^{\prime} d\phi^{\prime}  
W_{\xi \xi^{\prime} }(E_{\xi}, E_{\xi^{\prime}}) 
\Big[ 1 - \frac{ {\bf F}_{E_{\xi^{\prime}}} \cdot 
{\bf v}(E_{\xi^{\prime}},\theta^{\prime},\phi^{\prime}) }{ {\bf F}_{E_\xi} \cdot 
{\bf v}(E_\xi,\theta,\phi) } 
\frac{\tau_{\xi^{\prime}}(E_{\xi^{\prime}}) }{\tau_{\xi}(E_{\xi})} \Big]. 
\end{eqnarray}

First we consider $ E \geq 0$ case. After some straightforward calculation,
we get the following self-consistent equation for the relaxation time 
$\tau_{\lambda}(E)$:
\begin{eqnarray}
\frac{1}{\tau_{\lambda} (E)} = \sum_{\lambda^{\prime}}  
\frac{D_{\lambda^{\prime}}(E)}{2 u_0} 
\int d\theta^{\prime} \sin \theta^{\prime}(1+\lambda\lambda^{\prime} 
\cos\theta^{\prime})
\Big[ 1 -\cos\theta^{\prime}
\frac{\tau_{\lambda^{\prime}}(E)}{\tau_{\lambda}(E)} \Big].
\end{eqnarray}
Performing the integrals and summation, 
the above equation reduces to
\begin{eqnarray}
\frac{1}{\tau_{\pm}(E) } = \frac{2 D_{\pm}(E)}{3u_0} + \frac{D_{\mp}(E)}{u_0} 
\Big[ 1 + \frac{\tau_{\mp}( E) }{3 \tau_{\pm}(E)} \Big].
\end{eqnarray}
On solving the above coupled algebraic equations, the relaxation times 
of the two bands for $E \geq 0$ are obtained as 
\begin{eqnarray}
\tau_{\lambda}(E) = 
u_0 \Big[ \frac{D_{\lambda}(E)}{(D_{T}^{>}(E))^2} + \frac{1}{2D_{T}^{>}(E)} \Big].
\end{eqnarray}
For $ - E_{\alpha} < E <0$, the self-consistent equations for the relaxation time
$\tau_{\eta}(E)$ are 
\begin{eqnarray}
\frac{1}{\tau_{\eta} (E)} = \sum_{\eta^{\prime}}  \frac{D_{\eta^{\prime}}(E)}{2 u_0} 
\int d\theta^{\prime} \sin \theta^{\prime}(1+\cos\theta^{\prime})
\Big[ 1 -(-1)^{\eta^{\prime} - \eta}\cos\theta^{\prime}
\frac{\tau_{\eta^{\prime}}(E)}{\tau_{\eta}(E)} \Big].
\end{eqnarray}
The solutions for the relaxation times are obtained as
\begin{eqnarray}
\tau_{\eta}(E) = u_0 \Big[ \frac{D_{\eta}(E)}{(D_{T}^{<}(E))^2} 
+ \frac{1}{2D_{T}^{<}(E)} \Big] .
\end{eqnarray}
In our case, total density of states for $E\geq0$ and $E<0$ have 
the same form, 
\begin{eqnarray}
D_{T}^{>}(E) \equiv D_{T}^{<}(E) =  \frac{1}{2\pi^2} 
\Big(\frac{2m^\ast}{\hbar^2}\Big)^{\frac{3}{2}}
\Big[\frac{E+2E_\alpha}{\sqrt{E+E_\alpha} }\Big].
\end{eqnarray}
As a consequence of the same energy dependence of the total density of states 
below and above the BTP, we also have the same form of the relaxation
time for $E\geq0$,
\begin{align} \label{brta1}
\tau_{\lambda}(E) & = 2\pi^2 u_0 
\Big(\frac{\hbar^2}{2m^\ast}\Big)^{\frac{3}{2}} 
\Big(\frac{\sqrt{E + E_{\alpha}}}{E+2 E_{\alpha}}\Big) 
\Big[1 -  \lambda\frac{\sqrt{4E_{\alpha}(E + E_{\alpha})}}{(E + 2E_{\alpha})}\Big],
\end{align}
and for $E<0$,
\begin{align} \label{brta2}
\tau_{\eta}(E) & = 2\pi^2 u_0 \Big(\frac{\hbar^2}{2m^\ast}\Big)^{\frac{3}{2}} 
\Big(\frac{\sqrt{E + E_{\alpha}}}{E+2 E_{\alpha}}\Big) 
\Big[1-(-1)^{\eta-1}\frac{\sqrt{4E_{\alpha}(E+E_{\alpha})}}{(E+2E_{\alpha})}\Big].
\end{align}

\subsection{The electrical conductivity at $T=0$}
Within the semiclassical Boltzmann transport theory \cite{Ashcroft},
the general expression of the electrical conductivity at $T=0$ for $E\geq0$ is given by
\begin{eqnarray}
\sigma_{\nu \nu }^{>}=\frac{e^2}{4\pi}\sum_{\lambda=\pm 1} \int_{0}^{\infty} dE \;
(-\partial_{E}f^{0}(E) ) \int_{0}^{\pi}\int_{0}^{2\pi}\sin\theta d\theta d\phi 
D_{\lambda}(E) \langle \hat{v}_{\nu}(E,\theta,\phi)\rangle^2_{_{\lambda}} 
\tau_\lambda(E),
\end{eqnarray}
where $\nu =x,y,z $ and $\hat{v}_{\nu}(E,\theta,\phi)$ is the
$\nu$-component of the velocity operator and
$\tau_\lambda(E)$ is the relaxation time.
The expectation values of the velocity operator $\hat v_\mu $ with respect to the
$\lambda = \pm 1$ states are
$ \langle \hat{v}_{x}(E,\theta,\phi)\rangle_{ \lambda} =
\Big(\frac{ \hbar k_{\lambda}(E)}{m^\ast} + \frac{\lambda \alpha}{\hbar}\Big) 
\sin\theta\cos\phi$ and so on.
As we have already seen that $\tau_\lambda(E)$ is independent
of angular variables, so using
\begin{eqnarray}
\int_{0}^{\pi}\int_{0}^{2\pi}	\langle \hat{v}_{\nu}(E,\theta,\phi)\rangle^2_{_{\lambda}}
\sin\theta d\theta d\phi = \frac{4\pi}{3 }
\Big(\frac{\hbar k_{\lambda}(E)}{m^\ast} + \lambda \frac{\alpha}{\hbar}\Big)^2,
\end{eqnarray}
which exhibits isotropic nature of the electrical conductivity:
$ \sigma_{xx}=\sigma_{yy}=\sigma_{zz} = \sigma$.
Using the forms of density of states and Eq. (\ref{brta1}), we get
\begin{eqnarray} \label{cond1}
\sigma^> = \frac{e^2}{h} \frac{2 \hbar^2}{m^\ast n_{\rm imp} U_0^2}
\Big[ 1 - \frac{E_{F}^2}{3(2E_\alpha + E_F)^2} \Big]
(E_\alpha + E_F).
\end{eqnarray}

Similarly for $E<0$, the electrical conductivity becomes
\begin{eqnarray} \label{cond2}
\sigma^{<} & = &\frac{e^2}{4\pi}\sum_{\eta= 1,2} 
\int_{-E_{\alpha}}^{0} dE \;  (-\partial_{E}f^{0}(E))
\int_{0}^{\pi}\int_{0}^{2\pi}\sin\theta d\theta d\phi D_{\eta}(E)
\langle \hat{v}_{\nu}(E,\theta,\phi)\rangle^2_{_{\eta}} \tau_\eta(E), \nonumber\\
 & = & \frac{e^2}{h} \frac{2 \hbar^2}{m^\ast  n_{\rm imp} U_0^2}
\Big[ 1- \frac{E_{F}^2}{3(2E_\alpha + E_F)^2} \Big]
(E_\alpha + E_F).
\end{eqnarray}
It is clear from Eqs. (\ref{cond1}) and (\ref{cond2}) that electrical 
conductivity has same form below ($E_F<0$) and above $E_{F}\geq0$ the BTP. 
It is also clear from Fig. \ref{conductivities} that electrical conductivity 
has the same kind of energy dependence below and above the BTP.
So in our case electrical conductivity
will be continuously differentiable at the BTP.

\end{widetext}


\end{document}